\documentclass[
superscriptaddress,
aps,
twocolumn,
amsmath,
amssymb,
]{revtex4-1}

\usepackage{graphicx}
\usepackage{bm}
\usepackage{cases}
\usepackage{amsmath}

\newcommand{\ket}[1]{| #1\rangle}
\newcommand{\bra}[1]{\langle #1|}
\newcommand{\braket}[2]{\langle #1|#2\rangle}
\newcommand{\dket}[1]{| #1\rangle\!\rangle}
\newcommand{\dbra}[1]{\langle\!\langle #1|}
\newcommand{\dbraket}[2]{\langle\!\langle #1|#2\rangle\!\rangle}
\newcommand{\bigpar}[1]{\biggl( #1 \biggr)}

\newcommand{\hev}{{\rm hev.}}
\newcommand{\h}[2]{\{#1_{#2}\}}
\newcommand{\collop}[2]{\Psi^{(#1)}_{2}(#2)}

\begin{document}
\title{Microscopic Description of Quantum Lorentz Gas by Complex Spectral Representation of the Liouville-von Neumann Equation:\\ A Limitation of the Boltzmann Approximation}

\author{K. Hashimoto}
\email[Electric adress: ]{hashimotok@yamanashi.ac.jp}
\affiliation{
Graduate School of Interdisciplinary Research, University of Yamanashi, Kofu 400-8511, Japan
} 
\author{K. Kanki}
\author{S. Tanaka}
\affiliation{
Department of Physical Science, Osaka Prefecture University, Sakai 599-8531, Japan
}
\author{T. Petrosky}
\altaffiliation[Permanent address:]{
Center for Studies in Complex Quantum Systems, The University of Texas at Austin, TX 78712 USA
}
\affiliation{
Institute of Industrial Science, The University of Tokyo, Tokyo 153-8505, Japan}

\date{\today}

\begin{abstract}
Irreversible processes of one-dimensional quantum perfect Lorentz gas is studied 
on the basis of the fundamental laws of physics in terms of the complex spectral analysis 
associated with the resonance state of the Liouville-von Neumann operator. 
A limitation of the usual phenomenological Boltzmann equation is discussed 
from this dynamical point of view. 
For a Wigner distribution function that spreads over moderately small scale comparative to the mean-free-path, 
we found a shifting motion in space of the distribution that cannot be described 
by the hydrodynamic approximation of the kinetic equation. 
The mechanism of the shifting motion has two completely different origins: 
one is due to different value of the imaginary part of the eigenvalue of the Liouvillian 
and predominates in moderately small wavenumber associated to the spatial distribution, 
while the other is due to the existence of the real part of the eigenvalue associated to a wave propagation 
and predominates in moderately large wavenumber.

\begin{description}
\item[PACS number(s)]{05.30.-d, 05.60.-k, 05.20.Dd}
\end{description}
\end{abstract}

\pacs{Valid PACS appear here}

\maketitle

\section{introduction}

The purpose of this paper is to derive an irreversible kinetic equation through the fundamental laws of physics described by the Liouville-von Neumann equation without relying upon phenomenological arguments.  The example we discuss in this paper is a weakly-coupled one-dimensional (1D) quantum perfect Lorentz gas.

The world surrounding us is irreversible, and there is an arrow of time oriented toward our future that breaks time-symmetry. Nevertheless, the fundamental equations of motion in microscopic level are time-symmetric. To resolve this apparent contradiction is a long-standing problem in dynamics and statistical mechanics. A traditional assertion found in many textbooks in statistical mechanics is that the irreversibility is not a fundamental dynamical property, and it comes from our approximation of smearing out of detailed microscopic information that cannot be controlled due to our limitations as human beings (see e.g. \cite{Landau}). This anthropocentric view to explain the irreversibility is called {\it coarse-graining}. However, after we found the dissipative structure that allows our existence, this anthropocentric view is incomprehensible, since we are not fathers of the arrow of time, but we are children of the arrow of time \cite{Glansdorff1971,Nicolis1977}.

With Ilya Prigogine, one of the authors (TP) in this paper has formulated the irreversible dynamics through the fundamental laws of physics without relying upon such anthropocentric phenomenological arguments \cite{petpri97,Petrosky1996}. 
In this formulation we consider the spectral representation of the Liouville-von Neumann operator (Liouvillian) for systems in which irreversibility is expected such as the case of the thermodynamic limit where intensive variables and extensive variables exist. This formalism is an extension of the well-known Brillouin-Wigner-Feshbach formalism of the eigenvalue problem of the Hamiltonian in quantum systems to the eigenvalue problem of the Liouvillian. In this formulation, we start with a complete set of projection operators (often we choose a complete set of projection operators consisting of the eigenstates of the unperturbed Liouvillian).  Then, for each subspace associated to a projection operator we construct an {\it effective Liouvillian} that shares eigenvalues with the original Liouvillian, similarly to the case for the effective Hamiltonian in the Hamiltonian dynamics. 

Corresponding to the self-energy part of the effective Hamiltonian, we have {\it self-frequency part} of the effective Liouvillian. By the same reason in the effective Hamiltonian, the effective Liouvillian may become a non-Hermitian operator for unstable systems with resonance singularity, in spite of the fact that the Liouvillian itself is a Hermitian operator indicating the time-symmetric nature of the fundamental laws of physics. As a result, an eigenvalue of the Liouvillian becomes a complex number which is located at a resonance pole of the resolvent operator (or, the Green's function) of the Liouvillian in the complex frequency space. Associated to a complex eigenvalue, we have a resonance state for the Liouvillian as in the Hamiltonian case. For the Hamiltonian case, the imaginary part of a complex eigenvalue gives a decay rate of an unstable excited state. Similarly, the imaginary part of a complex eigenvalue of the Liouvillian gives a transport coefficient associated to an irreversible process. 

It is important to recognize that the existence of the resonance pole of the resolvent operator of the Hamiltonian is a direct consequence of the mathematical structure of the Hamiltonian, and does not result from an anthropocentric operation, such as the coarse-graining. In other words, the decay rate that breaks time-symmetry in the Hamiltonian system is an intrinsic dynamical property of the system. For exactly the same reason, irreversibility associated with transport coefficients in the Liouvillian system is an intrinsic dynamical property of the Liouvillian dynamics derived from the microscopic fundamental laws in physics.

In spite of the parallelism between the Hamiltonian case and the Liouvillian case, there is a fundamental difference between them in the self-energy or self-frequency part in the effective operators mentioned above. Indeed, even in the case where the unperturbed Hamiltonian does not degenerate, the unperturbed Liouvillian has intrinsic degeneracy.  For quantum case this is a result of the fact that the Liouvillian is defined by the commutation relation with the Hamiltonian (see (\ref{def:liouville})).
As a result, even in the case the self-energy part of the Hamiltonian is just a number, the self-frequency part becomes an operator. In statistical mechanics this self-frequency part operator has been called the {\it collision operator}, and it is a central object in non-equilibrium statistical physics.

Even though main framework of our formalism mentioned above of the irreversible processes is settled on the microscopic laws of the fundamental dynamics, its application to actual systems is generally extremely complex, because the infinitely many degrees of freedom are mutually interacting in systems that show irreversibility.
Moreover, the kinetic equation derived from the eigenvalue problem of the Liouvillian has a nonlinear structure that comes from two different origins; (i) one is due to the appearance of the product of the one-body distribution function due to the many-body effect, and (ii) the other is due to the fact that the collision operator itself depends on its eigenvalue.
Due to these complexity, we have applied this formulation so far only to the case where drastic approximations to solve the complex eigenvalue problem of the Liouvillian is applicable, such as the cases, for example, near equilibrium state where a linearized approximation with respect to the product of the distribution function is applicable, or the spatial inhomogeneity has a macroscopic scale at least of the order of the mean-free-length \cite{petrosky99-1,petrosky99-2,petrosky02}.
 Since one cannot analyze the dynamics in a microscopic scale much smaller than the mean-free-length by this drastic approximation, we have not yet been able to demonstrate the advantage of our microscopic approach over the phenomenological coarse-graining approach.

The main purpose of this paper (and successive forthcoming papers) is to analyze microscopic origin of irreversibility that is applicable to all spatial scale of the inhomogeneity including microscopic scale of the interaction range for the weekly-coupled 1D quantum perfect Lorentz gas where the mass of a light test particle is negligible as compared with mass of heavy particle.
The advantage of the Lorentz gas is a mathematical simplicity.
Thanks to the disparity in number of degrees of freedom between a light test particle and background many heavy particles, one can neglect the time evolution on the distribution function of the heavy particles in the thermodynamic limit. As a result, the complexity of  the nonlinearity associated to the origin (i) mentioned above does not exist in the Lorentz gas.

We especially focus our attention in this paper on limitations of the usual phenomenological Boltzmann approximation. Through our microscopic approach based on the complex spectral representation of the Liouvillian, the meaning of the coarse-graining will become clearer.

In addition, we report our discovery on an interesting shifting motion in space that is generated by the Boltzmann collision operator, especially for the distribution function localized in moderately small spatial scale less than the mean-free-length, but yet with a large enough width as compared with the microscopic scale given by the interaction range.
We found that there are two completely different mechanisms of the shifting motion in space; 
one is due to the asymmetry of the momentum distribution and predominates in a moderately small wavenumber associated to the inhomogeneity, while the other is due to the wave propagation associated to the real part in the eigenvalue that appears only in a moderately large wavenumber. As we shall see, this shifting motion cannot be described in the hydrodynamic approximation (such as the diffusion equation) that is applicable only to situations with extremely larger spatial scale as compared with the mean-free-length.

The structure of the present paper is as follows:
In Sec.II, we introduce the 1D quantum perfect Lorentz gas.
In Sec.III, we summarize general formalism of the complex spectral representation of the Liouvillian 
and apply it to the model.
In Sec.IV, we discuss limitations of the Boltzmann approximation.
In Sec.V, we present a solution of the eigenvalue problem of the Boltzmann collision operator and discuss its properties.
In Sec.VI, we discuss how the spectrum of the Boltzmann collision operator affects transport processes by analyzing time evolution of the Wigner distribution function.
In the last section VII, we give concluding remarks.
In Appendix, we show correction terms to the collision operator of the perfect Lorentz gas in a case where the ratio of the mass of the light particle to that of the heavy particles is not negligible.

\section{System}

We consider a weakly-coupled 1D quantum Lorentz gas.
The Lorentz gas consists of one light-mass particle (the test particle) with mass $m$ and $N$ heavy particles with mass $M$.
The Hamiltonian of the system is given by
	\begin{equation}
	H=H_{0}+gV=\frac{p^{2}}{2m}+\sum_{j=1}^{N}\frac{p_{j}^{2}}{2M}+g\sum_{j=1}^{N}V(|x-x_{j}|),
	\end{equation}
where $g$ is a dimensionless coupling constant 
and the interaction $V$ is assumed to be a short-range repulsive potential.
In this paper, we consider a weak-coupling situation ($g\ll1$).
In the following analysis, we restrict our attention to the limit
	\begin{equation}
	\frac{m}{M}\rightarrow0,
	\label{eq:perfect-lorentz-gas}
	\end{equation}
in which the system is called the {\it perfect Lorentz gas} \cite{Balescu63,Zhang95}.

We assume that the system is enclosed in a large 1D box of volume $L$ with the periodic boundary condition.
The interaction potential is expanded in the Fourier series as
	\begin{equation}
	V(|x-x_{j}|)=\frac{1}{\Omega}\sum_{n}V_{q_{n}}e^{iq_{n}(x-x_{j})},
	\label{eq:Fourier-potential}
	\end{equation}
where $V_{q_{n}}=V_{|q_{n}|}$, $\Omega\equiv L/2\pi$, the wavenumber $q_n$ is given by $q_{n}\equiv n\Delta q$ 
with $\Delta q\equiv1/\Omega$ and $n=0,\;\pm1,\;\pm2,\cdots$.
We assume $V_{q_{n}}$ is a continuous function at $q_{n}=0$ in the continuous limit $\Delta q \to 0$, 
and satisfies the condition,
\begin{equation}\label{eq:Vq_condition}
\operatorname{O}(|q_{n}|^{3/2})<\left|V_{q_{n}}\right|<\operatorname{O}(|q_{n}|^{1/2})
\end{equation}
for $q_{n} \to 0$, in order to avoid a singular transport process characteristic in 1D system 
(see (\ref{eq:gammaP}) for the decay rate, and (\ref{eq:diffusion-coeff}) for the diffusion coefficient).

In this paper, we consider the thermodynamic limit for the heavy particles,
	\begin{equation}
	L\rightarrow\infty,\hspace{10pt}N\rightarrow\infty,\hspace{10pt}c\equiv\frac{N}{L}={\rm finite},
	\label{eq:T-limit}
	\end{equation}
where $c$ is the concentration of the heavy particles.
In this limit, we have $\Delta q\rightarrow0$ and the wavenumber and the momentum become continuous variables.
At an appropriate stage, we shall replace a summation with an integration and a Kronecker delta $\delta^{Kr}$ with a Dirac $\delta$-function as
	\begin{equation}
	\frac{1}{\Omega}\sum_{q}\rightarrow\int dq,\hspace{10pt}\Omega_{\hbar}\delta^{Kr}(P-P')\rightarrow\delta(P-P'),
	\label{eq:continuous-limit}
	\end{equation}
with $\Omega_{\hbar}\equiv\Omega/\hbar$ (Hereafter we use a conventional notation $\sum_{q}$ for $\sum_{n}$ and drop the index $n$ in $q_{n}$).

Time evolution of the density matrix $\rho(t)$ obeys the Liouville-von Neumann equation,
	\begin{equation}
	i\frac{\partial}{\partial t}\rho(t)=L_{H}\rho(t),
	\end{equation}
where $L_{H}$ is the Liouvillian that is defined by the commutation relation with the Hamiltonian as
	\begin{equation}
	L_{H}\rho=\frac{1}{\hbar}[H,\rho]=(L_{0}+gL_{V})\rho,
	\label{def:liouville}
	\end{equation}
where $L_{0}$ is the unperturbed Liouvillian associated with $H_{0}$ and $gL_{V}$ is the interaction Liouvillian associated with $gV$.
In this paper, we focus our attention on the time evolution of the reduced density matrix for the test particle, which is defined by
	\begin{equation}
	f(t)\equiv{\rm Tr}_{{\rm hev}.}[\rho(t)],
	\label{eq:reduced-density-matrix}
	\end{equation}
where ${\rm Tr}_{{\rm hev}.}$ denotes the partial trace over the heavy particles.

We assume that the initial condition of the system is given in the form,
	\begin{equation}
	\rho(0)=f(0)\otimes\rho^{eq}_{{\rm hev}.},
	\end{equation}
where $\rho^{eq}_{{\rm hev}.}$ is the Maxwell distribution of the heavy particles with temperature $T$,
	\begin{equation}
        \left.
	\rho^{eq}_{{\rm hev}.}\equiv\prod^{N}_{j=1}\exp\left(-\frac{p^{2}_{j}}{2Mk_{B}T}\right)
        \!\!\right/\!\!
        {\rm Tr}\left[\exp\left(-\frac{p^{2}_{j}}{2Mk_{B}T}\right)\right],
	\label{eq:equilibriumdist}
	\end{equation}
and $k_{B}$ is the Boltzmann constant.
In the thermodynamic limit the time evolution of the density matrix associated with the heavy particles is negligible since its deviation from $\rho^{eq}_{{\rm hev}.}$ is proportional to $1/L$.

In order to discuss the ``coordinate'' and the ``momentum'' dependence of the distribution of the quantum particles 
in parallel with classical mechanics, 
let us introduce the Wigner distribution function, 
which is a quantum analog of the phase space distribution function in classical mechanics\cite{prigogine62,petpri97}, 
defined by
	\begin{eqnarray}
	&&\rho^{W}(X,\h{X}{j},P,\h{P}{j},t)\nonumber\\
	&&\equiv\frac{1}{\Omega^{N+1}}\!\!\sum_{k,\h{k}{j}}\!\rho_{k,\h{k}{j}}(P,\h{P}{j},t)e^{i(kX+k_{1}X_{1}+\cdots+k_{N}X_{N})},\nonumber\\
	\label{eq:10}
	\end{eqnarray}
with the Fourier component
	\begin{eqnarray}
	&&\rho_{k,\h{k}{j}}(P,\h{P}{j},t)\equiv\sqrt{\Omega_{\hbar}^{N+1}}\nonumber\\
	&&\times\biggr\langle P+\frac{\hbar k}{2},\biggr\{P_{j}+\frac{\hbar k_j}{2}\biggr\}\biggr|\rho(t)
\biggr|P-\frac{\hbar k}{2},\biggr\{P_{j}-\frac{\hbar k_j}{2}\biggr\}\biggr\rangle,\nonumber\\
	\label{def:wignerrepresentation}
	\end{eqnarray}
where the volume factor $\sqrt{\Omega_{\hbar}^{N+1}}$ in (\ref{def:wignerrepresentation}) comes from the normalization in the representation of the Wigner basis (see (\ref{eq:Wigner-basis})).
Here each of the notation $\h{X}{j}, \h{P}{j}$ represents a set of variables for the $N$ heavy particles,
and the momentum states of $N+1$ particles $|p,\{p_{j}\}\rangle$
is an eigenvector of $H_{0}$,
	\begin{equation}
	H_{0}\ket{p,\{p_{j}\}}=\epsilon_{p,\{p_{j}\}}\ket{p,\{p_{j}\}},
	\label{eq:app:H0}
	\end{equation}
with
	\begin{equation}
	\epsilon_{p,\{p_{j}\}}\equiv\frac{p^{2}}{2m}+\sum_{j=1}^{N}\frac{p_{j}^{2}}{2M}.
	\end{equation}
The ``wavenumber'' $k$ and the ``momentum'' $P$ in the Fourier component of the Wigner representation 
are given in terms of the momenta in the matrix elements $\langle p,\{p_{j}\}|\rho(t)|p',\{p_{j}'\}\rangle$ as 
	\begin{eqnarray}
	\begin{split}
	&\hbar k=p-p',\hspace{10pt}P=\frac{1}{2}(p+p'),\\
	&\hbar k_{j}=p_{j}-p_{j}',\hspace{10pt}P_{j}=\frac{1}{2}(p_{j}+p_{j}').
	\label{eq:wignervariable}
	\end{split}
	\end{eqnarray}

\section{Complex spectral representation of the Liouvillian}

\subsection{Preliminaries}
In this subsection, we give a brief summary of the Liouville space representation of the Hilbert space operators. (see \cite{petpri97} for more detail)

The Liouville space is spanned by linear operators $A$, $B$, $\cdots$ in the ordinary wave function space.
We represent these linear operators by double bra-ket notations as $\dket{A}$, $\dket{B}$,$\cdots$ in order to distinguish them from the bra-ket notation in the wave function space.
The inner product in the Liouville space is defined by
	\begin{equation}
	\dbraket{B}{A}\equiv{\rm Tr}[B^{+}A],
	\label{eq:innerprod}
	\end{equation}
where $B^{+}$ is the Hermitian conjugate of a linear operator $B$ in the wave function space.

In the Liouville space, we can consider a linear operator ${\cal T}$ acting on vectors, 
which is operators in the wave function space.
We call it a super-operator in case we want to emphasize the difference between it from the linear operator in the wave function space.
One can define a super-operator ${\cal T}^{\dagger}$ that is the Hermitian conjugate to ${\cal T}$ as
	\begin{equation}
	\dbra{A}{\cal T}^{\dagger}\dket{B}\equiv (\dbra{B}{\cal T}\dket{A})^{c.c.},
	\end{equation}
where the notation $c.c.$ means the complex conjugate.
Then, one can introduce the Hermitian super-operator which satisfies
	\begin{equation}
	{\cal T}^{\dagger}={\cal T},
	\end{equation}
and the unitary super-operator which satisfies
	\begin{equation}
	U^{\dagger}=U^{-1}.
	\end{equation}
The Liouvillian is an example of the Hermitian super-operators in the Liouville space, i.e.,
	\begin{equation}
	L_{H}^{\dagger}=L_{H},
	\end{equation}
and the time evolution operator
	\begin{equation}
	{\cal U}(t)=e^{-iL_{H}t},
	\end{equation}
is an example of the unitary super-operators.

We now introduce a basis set in the Liouville space for 1D quantum Lorentz gas in terms of the eigenstates of $H_{0}$ in (\ref{eq:app:H0}).
Let us introduce an abbreviated notation of the $N+1$ variables ${\bf p}=\{p,\;p_{1},\cdots,\;p_{N}\}$ to avoid heavy notations.
Then, we express the eigenstate and the eigenvalue of $H_{0}$ in (\ref{eq:app:H0}) as $\ket{{\bf p}}$ and $\epsilon_{{\bf p}}$.
The eigenstates of $H_{0}$ satisfy the orthonormality relation
	\begin{equation}
	\braket{{\bf p}}{{\bf p}'}=\delta^{Kr}({\bf p}-{\bf p}')\equiv\delta^{Kr}(p-p')\prod_{j=1}^{N}\delta^{Kr}(p_{j}-p'_{j}),
	\end{equation}
and the completeness relation
	\begin{equation}
	\sum_{{\bf p}}\ket{{\bf p}}\!\bra{{\bf p}}=1.
	\end{equation}

We define vectors in the Liouville space by
	\begin{equation}
	\dket{{\bf p},{\bf p}'}\equiv\ket{{\bf p}}\!\bra{{\bf p}'}.
	\end{equation}
These vectors are eigenstates of $L_{0}$,
	\begin{equation}\label{eq:unperturbed_eigenstates}
	L_{0}\dket{{\bf p},{\bf p}'}=\frac{1}{\hbar}(\epsilon_{{\bf p}}-\epsilon_{{\bf p}'})\dket{{\bf p},{\bf p}'}.
	\end{equation}
The set of eigenstates $\{\dket{{\bf p},{\bf p}'}\}$ forms a orthonormal basis in the Liouville space (see (\ref{eq:innerprod}))
	\begin{equation}
	\dbraket{{\bf p},{\bf p}'}{{\bf p}'',{\bf p}'''}=\delta^{Kr}({\bf p}-{\bf p}'')\delta^{Kr}({\bf p}'-{\bf p}'''),
	\label{eq:normalization-p}
	\end{equation}
and the completeness relation
	\begin{equation}
	\sum_{{\bf p}}\sum_{{\bf p}'}\dket{{\bf p},{\bf p'}}\!\dbra{{\bf p},{\bf p}'}=1.
	\end{equation}

We note that the eigenvalues of $L_0$ in (\ref{eq:unperturbed_eigenstates}) 
are degenerated (e.g. for ${\bf p}'={\bf p}$) 
even though $\epsilon_{\bf p}$ are not degenerated. 
This is the intrinsic degeneracy of the Liouvillian mentioned in the introduction.

In terms of the basis, we have a simple expression for a matrix element of an operator $A$ in the wave function space as
	\begin{equation}
	A_{{\bf p},{\bf p}'}\equiv\bra{{\bf p}}A\ket{{\bf p}'}=\dbraket{{\bf p},{\bf p}'}{A}.
	\end{equation}
A matrix element of the interaction Liouvillian is given by
	\begin{eqnarray}
	\begin{split}
	&\dbra{{\bf p},{\bf p}'}gL_{V}\dket{{\bf p}'',{\bf p}'''}\\
	&=\frac{1}{\hbar}[gV_{{\bf p},{\bf p}''}\delta^{Kr}({\bf p}'-{\bf p}''')-\delta^{Kr}({\bf p}-{\bf p}'')gV_{{\bf p}''',{\bf p}'}].
	\label{app:mat_int}
	\end{split}
	\end{eqnarray}

We define the Wigner basis by
	\begin{equation}
	\dket{{\bf k};{\bf P}}\equiv\sqrt{\Omega^{N+1}_{\hbar}}\dket{{\bf p},{\bf p'}},
	\label{eq:Wigner-basis}
	\end{equation}
with the relations in (\ref{eq:wignervariable}), where the volume factor $\sqrt{\Omega^{N+1}_{\hbar}}$ is introduced to take a continuous limit $\Omega_{\hbar}\rightarrow\infty$ on the variables $P$, $\{P_{j}\}$ (see (\ref{eq:continuous-limit}) and (\ref{eq:orthonormal-Wigner})) analogous to the classical system where momenta are intrinsically continuous variables \cite{Petrosky1996}.
The set of the Wigner basis also forms a complete orthonormal basis
	\begin{eqnarray}
	&&\dbraket{{\bf k};{\bf P}}{{\bf k}';{\bf P}'}=\!\delta^{Kr}(k-k')\delta_{\Omega_{\hbar}}(P-P')\nonumber\\
	&&\hspace{65pt}\times\prod_{j=1}^{N}\delta^{Kr}(k_{j}-k'_{j})\delta_{\Omega_{\hbar}}(P_{j}-P'_{j}),
	\label{eq:orthonormal-Wigner}
	\end{eqnarray}
with
	\begin{equation}
	\delta_{\Omega_{\hbar}}(P-P')\equiv\Omega_{\hbar}\delta^{Kr}(P-P'),
	\end{equation}
and
	\begin{equation}
	\sum_{{\bf k}}\frac{1}{\Omega_{\hbar}^{N}}\sum_{{\bf P}}\dket{{\bf k};{\bf P}}\!\dbra{{\bf k};{\bf P}}=1.
	\end{equation}
With the Fourier component of the interaction potential in (\ref{eq:Fourier-potential}), the Wigner representation of the matrix element of the interaction Liouvillian (\ref{app:mat_int}) is given by
	\begin{eqnarray}
	&&\dbra{{\bf k};{\bf P}}gL_{V}\dket{{\bf k}';{\bf P}'}\nonumber\\
	&&=\frac{1}{\Omega}\sum_{j=1}^{N}\!\frac{gV_{k-k'}}{\hbar}\delta^{Kr}(k-k'+k_{j}-k_{j}')\!\!\prod_{i(\not=j)}^{N-1}\!\!\delta^{Kr}(k_{i}-k_{i}')\nonumber\\
	&&\hspace{10pt}\times[{\hat\eta}_{P}^{\frac{\hbar}{2}(k-k')}{\hat\eta}_{P_{j}}^{-\frac{\hbar}{2}(k-k')}-{\hat\eta}_{P}^{-\frac{\hbar}{2}(k-k')}{\hat\eta}_{P_{j}}^{\frac{\hbar}{2}(k-k')}]\nonumber\\
	&&\hspace{10pt}\times\delta_{\Omega_{\hbar}}({\bf P}-{\bf P}'),
	\label{appeq:matrix_interaction}
	\end{eqnarray}
with $\hbar{\bf k}'\equiv{\bf p}''-{\bf p}'''$ and ${\bf P}'\equiv({\bf p}''+{\bf p}''')/2$, where ${\hat\eta}_{P}^{p}$ and ${\hat\eta}_{P_{j}}^{p_{j}}$ are displacement operators acting on the momenta $P$ and $P_{j}$, respectively, as
	\begin{equation}
	{\hat\eta}_{P}^{p}f(P)=f(P+p),\hspace{10pt}{\hat\eta}_{P_{j}}^{p_{j}}f(P_{j})=f(P_{j}+p_{j}).
	\label{eq:displacement}
	\end{equation}

\subsection{General formalism}

The eigenvalue problem of the Liouvillian is given by
	\begin{subequations}
	\begin{eqnarray}
	&&L_{H}\dket{F^{(\nu)}_{\alpha}}=Z^{(\nu)}_{\alpha}\dket{F^{(\nu)}_{\alpha}},\label{eq:eigenvalue-eq-r}\\
	&&\dbra{{\tilde F}^{(\nu)}_{\alpha}}L_{H}=\dbra{{\tilde F}^{(\nu)}_{\alpha}}Z^{(\nu)}_{\alpha},\label{eq:eigenvalue-eq-l}
	\end{eqnarray}
	\end{subequations}
where the indices $\alpha$ and $\nu$ specify an eigenvalue 
(especially $\nu$ denotes the spatial correlation subspace (see \cite{petpri97})), and
$\dket{F^{(\nu)}_{\alpha}}$ and $\dbra{{\tilde F}^{(\nu)}_{\alpha}}$ are right- and left-eigenstates, respectively.

We introduce effective Liouvillians 
by using the Brillouin-Wigner-Feshbach formalism \cite{petpri97,petrosky2010} 
with projection operators $P^{(\nu)}$ and $Q^{(\nu)}$ satisfying
	\begin{equation}
	P^{(\nu)}L_{0}=L_{0}P^{(\nu)},
	\end{equation}
	\begin{equation}
	P^{(\nu)}P^{(\mu)}=\delta_{\nu,\mu},
	\end{equation}
	\begin{equation}
	\sum_{\nu}P^{(\nu)}={\hat I}_{N+1},
	\end{equation}
and
	\begin{equation}
	P^{(\nu)}+Q^{(\nu)}={\hat I}_{N+1},
	\end{equation}
where ${\hat I}_{N+1}$ is the unit operator for the $N+1$ particle system.
By applying these projection operators on (\ref{eq:eigenvalue-eq-r}), the eigenvalue equation of the Liouvillian takes the form
	\begin{equation}
	\Psi^{(\nu)}(Z^{(\nu)}_{\alpha})P^{(\nu)}\dket{F^{(\nu)}_{\alpha}}=Z^{(\nu)}_{\alpha}P^{(\nu)}\dket{F^{(\nu)}_{\alpha}},
	\label{eq:dispersion}
	\end{equation}
where
	\begin{eqnarray}
	\begin{split}
	&\Psi^{(\nu)}(z)\equiv P^{(\nu)}L_{H}P^{(\nu)}\\
	&+P^{(\nu)}L_{H}Q^{(\nu)}\frac{1}{z-Q^{(\nu)}L_{H}Q^{(\nu)}}Q^{(\nu)}L_{H}P^{(\nu)}
	\label{eq:collop}
	\end{split}
	\end{eqnarray}
is the {\it effective Liouvillian}. Its second term is the self-frequency part 
that corresponds to the self-energy part of an effective Hamiltonian in the case of the Hamiltonian operator
in the wave function space.
The effective Liouvillian is also called the {\it collision operator} which is a central object 
in the kinetic theory in non-equilibrium statistical mechanics \cite{prigogine62,petpri97}.
One can see from its eigenvalue equation (\ref{eq:dispersion}) that the collision operator 
shares the eigenvalues with the Liouvillian.
The eigenvalue equation of the collision operator (\ref{eq:dispersion}) is {\it non-linear}, 
i.e. the collision operator itself depends on the eigenvalue.

In terms of the right- and left-eigenstates of the collision operator $\Psi^{(\nu)}(z)$, the right- and the left-eigenvectors of the Liouvillian $L_{H}$ are given by
	\begin{subequations}
	\begin{eqnarray}
	&&\dket{F^{(\nu)}_{\alpha}}=\bigr[P^{(\nu)}+{\cal C}^{(\nu)}(Z^{(\nu)}_{\alpha})\bigr]P^{(\nu)}\dket{F^{(\nu)}_{\alpha}},\\
	&&\dbra{{\tilde F}^{(\nu)}_{\alpha}}=\dbra{{\tilde F}^{(\nu)}_{\alpha}}P^{(\nu)}\bigr[P^{(\nu)}+{\cal D}^{(\nu)}(Z^{(\nu)}_{\alpha})\bigr],
	\end{eqnarray}
	\end{subequations}
with the {\it creation-of-correlation operator}
	\begin{subequations}
	\begin{equation}
	{\cal C}^{(\nu)}(z)=\frac{1}{z-Q^{(\nu)}L_{H}Q^{(\nu)}}Q^{(\nu)}L_{H}P^{(\nu)},
	\end{equation}
and the {\it destruction-of-correlation operator}
	\begin{equation}
	{\cal D}^{(\nu)}(z)=P^{(\nu)}L_{H}Q^{(\nu)}\frac{1}{z-Q^{(\nu)}L_{H}Q^{(\nu)}},
	\end{equation}
	\end{subequations}
which are off-diagonal transitions between the $Q^{(\nu)}$ subspace and the $P^{(\nu)}$ subspace \cite{petpri97}.

It is well-known for an unstable quantum system with a continuous spectrum that the effective Hamiltonian becomes a non-Hermitian operator due to the resonance singularity in the self-energy part \cite{Hatano13}.
Similarly, the collision operator becomes a non-Hermitian operator in the Liouville space in the thermodynamic limit.
As a result, the collision operator has eigenstates with complex eigenvalues that are called {\it resonance states}.
The imaginary part of the complex eigenvalue of the Liouvillian 
gives a transport coefficient of the system \cite{petrosky2010} 
(see also (\ref{eq:expansion_hydrodynamics}) and (\ref{eq:diffusion-coeff})).

\subsection{Application to the weakly-coupled 1D quantum perfect Lorentz gas}

Let us now apply the general formalism presented above 
to the weakly-coupled 1D quantum perfect Lorentz gas.
Using the Wigner basis (\ref{eq:Wigner-basis}), the Fourier component of the Wigner distribution function (\ref{def:wignerrepresentation}) is represented by
	\begin{equation}
	\rho_{k,\h{k}{j}}(P,\h{P}{j},t)=\dbraket{k,\h{k}{j};P,\h{P}{j}}{\rho(t)}.
	\end{equation}

In order to apply the Brilloun-Wigner-Feshbach formalism, 
we define the projection operators as
	\begin{equation}
	P^{(k)}\!\equiv\!\frac{1}{\Omega_{\hbar}^{N+1}}\sum_{{\bf P}}\!\dket{k,\{0_{j}\};{\bf P}}\!\dbra{k,\{0_{j}\};{\bf P}},
	\label{eq:projectionop}
	\end{equation}
and
	\begin{equation}
	Q^{(k)}=1-P^{(k)}=\!\!\!\!\sum_{{\tilde{\bf k}}\not=(k,\{0\})}\frac{1}{\Omega^{N+1}_{\hbar}}\sum_{{\tilde{\bf P}}}\dket{{\tilde{\bf k}};{\tilde{\bf P}}}\!\dbra{{\tilde{\bf k}};{\tilde{\bf P}}},
	\end{equation}
where we have used the notation $\{0_{j}\}$ to indicate that all wavenumbers associated to the heavy particles are zero.
For the projection operator $P^{(k)}$, we have
	\begin{equation}
	P^{(k)}L_{0}P^{(k)}\dket{k,\{0_{j}\};{\bf P}}=\frac{kP}{m}\dket{k,\{0_{j}\};{\bf P}},
	\label{eq:flow-term}
	\end{equation}
and
	\begin{equation}
	gP^{(k)}L_{V}P^{(k)}=0,
	\label{eq:conditionforP}
	\end{equation}
because of $V_0=0$ due to the assumption (\ref{eq:Vq_condition}).

In the weak coupling situation, the collision operator (\ref{eq:collop}) can be approximated 
up to the second order in $g$ as
	\begin{eqnarray}
	\begin{split}
	\Psi^{(k)}_{2}(z)&=P^{(k)}L_{0}P^{(k)}\\
	&\;\;+g^{2}P^{(k)}L_{V}Q^{(k)}\frac{1}{z-L_{0}}Q^{(k)}L_{V}P^{(k)}.
	\label{eq:collop-weak}
	\end{split}
	\end{eqnarray}
Using (\ref{appeq:matrix_interaction}) and  (\ref{eq:flow-term}), we have
	\begin{eqnarray}
	&&\dbra{k,\{0_{j}\};{\bf P}}\collop{k}{z}\dket{k,\{0_{j}\};{\bf P}'}\nonumber\\
	&&=\biggl[\frac{kP}{m}-\frac{1}{\Omega^2}\sum_{j=1}^{N}\sum_{q}\frac{g^2|V_{q}|^2}{\hbar^2}({\hat\eta}^{\frac{\hbar}{2}q}_{P}{\hat\eta}^{-\frac{\hbar}{2}q}_{P_j}-{\hat\eta}^{-\frac{\hbar}{2}q}_{P}{\hat\eta}^{\frac{\hbar}{2}q}_{P_j})\nonumber\\
	&&\hspace{5pt}\times\frac{1}{z-(k-q)P/m-qP_j/M}({\hat\eta}^{\frac{\hbar}{2}q}_{P}{\hat\eta}^{-\frac{\hbar}{2}q}_{P_j}-{\hat\eta}^{-\frac{\hbar}{2}q}_{P}{\hat\eta}^{\frac{\hbar}{2}q}_{P_j})\biggr]\nonumber\\
	&&\hspace{5pt}\times\delta_{\Omega_{\hbar}}({\bf P}-{\bf P}').
	\label{app:collop-weak}
	\end{eqnarray}

We focus our attention on the test particle.
We define the reduced collision operator for the test particle as
	\begin{equation}
	{\bar\Psi}^{(k)}_{2}(z)\equiv{\rm Tr}_{{\rm hev}.}[\Psi^{(k)}_{2}(z)\rho^{eq}_{{\rm hev}.}],
	\end{equation}
where we put a bar on the notation to distinguish it 
from the collision operator for the whole system (\ref{eq:collop-weak}).
Let us now derive an expression of its matrix element,
	\begin{eqnarray}
	&&\!\dbra{k;P}{\bar\Psi}^{(k)}_{2}(z)\dket{k;P'}\nonumber\\
	&&=\dbra{k;P}\biggr[\sum_{\{P_{j}\}}\frac{1}{\Omega_{\hbar}^{N}}\dbra{\{0_{j}\};\{P_{j}\}}\Psi^{(k)}_{2}(z)\dket{\rho^{eq}_{{\rm hev}.}}\biggr]\dket{k;P'}.\nonumber\\
	\end{eqnarray}
	Taking a large volume limit $L\rightarrow\infty$ for the momentum variables and performing the integration over the momenta with the aid of (\ref{eq:equilibriumdist}) and (\ref{app:collop-weak}), we have
	\begin{eqnarray}
	&&\dbra{k;P}{\bar\Psi}^{(k)}_{2}(z)\dket{k;P'}=\!\!\biggl[\frac{kP}{m}-g^{2}\frac{1}{\Omega^{2}}\sum_{j=1}^{N}\int^{\infty}_{-\infty}dP_{j}\nonumber\\
	&&\times\sum_{q\not=0}\frac{|V_{q}|^{2}}{\hbar^{2}}({\hat\eta}^{\frac{\hbar}{2}q}_{P}{\hat\eta}^{-\frac{\hbar}{2}q}_{P_{j}}-{\hat\eta}^{-\frac{\hbar}{2}q}_{P}{\hat\eta}^{\frac{\hbar}{2}q}_{P_{j}})\nonumber\\
	&&\times\frac{1}{z-(k-q)P/m-lP_j/M}({\hat\eta}^{\frac{\hbar}{2}q}_{P}{\hat\eta}^{-\frac{\hbar}{2}q}_{P_{j}}-{\hat\eta}^{-\frac{\hbar}{2}q}_{P}{\hat\eta}^{\frac{\hbar}{2}q}_{P_{j}})\nonumber\\
	&&\times\rho^{eq}_{\hev}(P_{j})\biggr]\delta_{\Omega_{\hbar}}(P-P'),
	\label{eq:matrix-element-1}
	\end{eqnarray}
where $\rho^{eq}_{\hev}(P_{j})$ is the Maxwell distribution for the heavy particle $j$ with temperature $T$,
	\begin{equation}
	\rho^{eq}_{\hev}(P_{j})\equiv\biggr(\frac{1}{2\pi Mk_{B}T}\biggr)^{1/2}\exp\biggr(-\frac{P^{2}_{j}}{2Mk_{B}T}\biggr).
	\end{equation}
The first and second terms in (\ref{eq:matrix-element-1}) are called the ``flow term'' and the ``collision term'', 
respectively.

Since the mass of the test particle $m$ is much smaller than the mass of a heavy particle $M$, 
we expand the propagator in (\ref{eq:matrix-element-1}) as a power series of the ratio $m/M$.
Then we have
	\begin{equation}
	\dbra{k;P}{\bar\Psi}^{(k)}_{2}(z)\dket{k;P'}=\dbra{k;P}\psi^{(k)}(z)\dket{k;P'}+O\biggr(\frac{m}{M}\biggr),
	\label{eq:mass-exp}
	\end{equation}
where
	\begin{eqnarray}
	&&\dbra{k;P}\psi^{(k)}(z)\dket{k;P'}=\biggr[\frac{kP}{m}-\frac{2\pi g^{2}c}{\hbar^{2}}\frac{1}{\Omega}\sum_{q\not=0}|V_{q}|^{2}\nonumber\\
	&&\times\partial^{\hbar q/2}_{P}\frac{1}{z-(k-q)P/m}\partial^{\hbar q/2}_{P}\biggr]\delta_{\Omega_{\hbar}}(P-P'),
	\label{eq:collop-pl}
	\end{eqnarray}
is the collision operator for the {\it perfect} Lorentz gas ($m/M\rightarrow0$).
Here $\partial^{\hbar q/2}_{P}$ is defined by
	\begin{equation}
	\partial^{\hbar q/2}_{P}\equiv{\hat\eta}^{\frac{\hbar}{2}q}_{P}-{\hat\eta}^{-\frac{\hbar}{2}q}_{P}.
	\end{equation}
Note that there is no temperature dependent term in (\ref{eq:collop-pl}).
Temperature dependence comes from the correction term to (\ref{eq:collop-pl}) 
starting with the first order in $(m/M)$ (see Appendix).

For the reduced collision operator, we write the eigenvalue problem as
	\begin{subequations}
	\begin{equation}
	\psi^{(k)}(z^{(k)}_{\alpha})\dket{u^{(k)}_{\alpha}}=z^{(k)}_{\alpha}\dket{u^{(k)}_{\alpha}},
	\end{equation}
	\begin{equation}
	\dbra{{\tilde v}^{(k)}_{\alpha}}\psi^{(k)}(z^{(k)}_{\alpha})=z^{(k)}_{\alpha}\dbra{{\tilde v}^{(k)}_{\alpha}}.
	\end{equation}
	\label{eq:main}
	\end{subequations}
We note that $z^{(k)}_{\alpha}=Z^{(k)}_{\alpha}$ for our Lorentz gas,
because the heavy particles are in an eigenstate with zero eigenvalue,
i.e. they remain in thermal equilibrium.

The collision operator in (\ref{eq:main}) depends on its eigenvalue.
In this sense, the eigenvalue equation is still nonlinear.
Our main goal is to construct the solution of the nonlinear eigenvalue problem. 
As a preparation to achieve this goal, however, we here restrict our attention on the linear approximation. 
The complete basis obtained for the linear problem presented in this paper will be used to achieve the main goal in a forthcoming paper.

\section{The Boltzmann collision operator}\label{sec:bolts}

To solve the eigenvalue problem (\ref{eq:main}), we here study a situation where the wavenumber $k$ satisfies
	\begin{equation}
	|k|\ll d^{-1},
	\label{eq:bol-region}
	\end{equation}
where $d$ is the interaction range between the particles.
In this situation in addition to the weak-coupling, 
we shall show that the eigenvalue dependence of the collision operator (\ref{eq:collop-pl}) 
is negligible and the collision operator $\psi^{(k)}(z)$ is reduced 
to the phenomenological Boltzmann collision operator.

For a spatial inhomogeneity satisfying the condition (\ref{eq:bol-region}), 
a typical value of $q$ appearing in (\ref{eq:Fourier-potential}) is much larger than $k$ in (\ref{eq:collop-pl}),
	\begin{equation}
	|k|\ll|q|.
	\label{eq:kllq}
	\end{equation}
Then we can neglect $k$ in the denominator in the second term in (\ref{eq:collop-pl}).
We note that since the collision term in (\ref{eq:collop-pl}) is an operator acting on the momentum $P$, 
this term does not commute with the flow term. 
This property changes the $k$-dependence of the eigenvalue of $\psi^{(k)}$ from the linear $k$-dependence 
in the flow term to some different $k$-dependence. 
Indeed, the component of the eigenvalue associated to the diffusion processes starts
with $k^2$ in the series expansion in $k$ (see (\ref{eq:expansion_hydrodynamics})).

On the other hand, we may expect that the imaginary part of the eigenvalue $z$ in (\ref{eq:main}) 
is proportional to $g^2$ for $ g \ll 1$ because of the factor $g^2$ in front of the collision term in (\ref{eq:collop-pl}). 
If this is the case, we can replace $z$ in (\ref{eq:collop-pl}) by $+i0$,
	\begin{equation}
	\psi^{(k)}(z^{(k)}_{\alpha})=\psi^{(k)}(+i0)+O(g^{4}).
	\end{equation}
Here $+i0$ means that the collision operator $\psi^{(k)}(z)$ is evaluated on the real axis approaching from the upper half-plane of $z$ to ensure the time evolution is oriented to the future $t>0$ \cite{petpri97}.
Combining these arguments, we can approximate  $\psi^{(k)}(z_\alpha^{(k)})$ by the new collision operator given by
	\begin{eqnarray}
	&&\dbra{k;P}\psi^{(k)}_{B}\dket{k;P'}\equiv\biggr[\frac{kP}{m}-\frac{2\pi g^{2}c}{\hbar^{2}}\nonumber\\
	&&\times\lim_{\epsilon\rightarrow+0}\!\int^{\infty}_{-\infty}\!\!\!dq|V_{q}|^{2}\partial^{\hbar q/2}_{P}\frac{1}{+i\epsilon+qP/m}\partial^{\hbar q/2}_{P}\biggr]\nonumber\\
	&&\times\delta(P-P'),
	\label{eq:collop-matel}
	\end{eqnarray}
where we have taken the thermodynamic limit (\ref{eq:T-limit}).
This is the phenomenological Boltzmann collision operator for the 1D quantum perfect Lorentz gas. \cite{Zhang95}

We note that in spite of the fact that $\psi^{(k)}_B$ is in the $P^{(k)}$ subspace 
that is orthogonal to the $P^{(0)}$ subspace, the second term of 
the matrix element of $\psi^{(k)}_B$ does not depend on $k$ and is the same as 
the matrix element of the $\psi^{(0)}_B$ that is associated to the spatially homogeneous component $k=0$ and generates a time evolution of the momentum distribution function. 
In this sense the phenomenological Boltzmann collision operator is a ``coarse-grained object'' in space, 
and it does not contain information of microscopic structure of the interaction.  
As a result, the Boltzmann collision operator is not applicable to a situation with so small scale 
of spatial structure that the microscopic shape of the interaction starts to play a role.

We also note that the second term in (\ref{eq:collop-matel}) vanishes for the classical 1D perfect Lorentz gas.
This is because the momentum distribution function cannot change in time by an elastic two-body collision in the classical 1D system (for the perfect Lorentz gas, or for the system that consists of same mass particles).
This is not the case in quantum system, because there is a forward scatting due to a quantum effect in addition to a backward scattering.
In this sense, the quantum system has an advantage on the mathematical simplicity over the classical corresponding systems.

Furthermore, we note that the above derivation of  the Boltzmann collision operator is still semi-quantitative, because the applicability of this approximation as (\ref{eq:bol-region}) is too rough.
Precise limitation of the Boltzmann approximation will be clarified in the forthcoming paper by solving the nonlinear problem of the collision operator with the aid of the complete eigen-basis of the Boltzmann collision operator obtained in this paper.

By performing the $q$ integration in (\ref{eq:collop-matel}) with a formula
	\begin{equation}
	\lim_{\epsilon\rightarrow+0}\frac{1}{x+i\epsilon}=-i\pi\delta(x)+{\cal P}\frac{1}{x},
	\end{equation}
where ${\cal P}$ denotes the principal part, a matrix element of the collision operator is expressed as
	\begin{eqnarray}
	&&\dbra{k;P}\psi^{(k)}_{B}\dket{k;P'}=\frac{kP}{m}\delta(P-P')\nonumber\\
	&&\hspace{50pt}+i\frac{g^2\gamma_{P}}{2}[\delta(P+P')-\delta(P-P')],
	\label{eq:matrixel-boltzmann}
	\end{eqnarray}
with
	\begin{equation}\label{eq:gammaP}
	\gamma_{P}\equiv \frac{8\pi^{2}mc}{\hbar^{2}|P|}|V_{\frac{2P}{\hbar}}|^{2}.
	\end{equation}
We have $\gamma_P\to 0$ for $P\to 0$ (see (\ref{eq:Vq_condition})).

From the expression (\ref{eq:matrixel-boltzmann}), one can see that the Boltzmann collision operator has non-vanishing matrix elements only between the states $\dket{k;P}$ and $\dket{k;-P}$.
Physically, this is because there are only forward and backward scattering in the 1D quantum system.
Hence, in terms of this basis, the Boltzmann collision operator is represented by a $2\times2$ non-Hermitian matrix,
	\begin{equation}
	\psi^{(k)}_{B}=
	\begin{pmatrix}
	kP/m-ig^2\gamma_{P}/2 & ig^2\gamma_{P}/2 \\
	ig^2\gamma_{P}/2 & -kP/m-ig^2\gamma_{P}/2 \\
	\end{pmatrix}.
	\label{eq:2matrix-boltzmann}
	\end{equation}

In terms of the operator $\psi^{(k)}_{B}$, the time evolution equation for the reduced density matrix 
for the test particle is approximated by
	\begin{equation}
	i\frac{\partial}{\partial t}{\hat p}^{(k)}\dket{f(t)}=\psi^{(k)}_{B}{\hat p}^{(k)}\dket{f(t)},
	\label{eq:bolt-eq}
	\end{equation}
where
	\begin{equation}
	{\hat p}^{(k)}\equiv\frac{1}{\Omega_{\hbar}}\sum_{P}\dket{k;P}\!\dbra{k;P}.
	\end{equation}


\section{Eigenstates of the Boltzmann collision operator}

In this section, we present the eigenstates of the Boltzmann collision operator (\ref{eq:2matrix-boltzmann}) 
and summarize their properties.
The structure of the collision operator (\ref{eq:2matrix-boltzmann}) is so simple 
that the solution of its eigenvalue problem itself has already been known \cite{Zhang95}.  
We present here the explicit form of the solution in order to show a new physical insight 
into the time evolution generated by the Boltzmann collision operator, 
especially for the distribution function localized in moderately small spatial region of a scale 
less than the mean-free-length, 
but yet with a large enough width as compared with the microscopic scale given by the interaction range.
In the following, we shall separately treat situations $k=0$ and $k\not=0$.

\subsection{The spatially homogeneous situation: $k=0$}

We first consider the spatially homogenous situation in the $k=0$ subspace.
This subspace corresponds to momentum distribution function.
Then, the Boltzmann collision operator is given by an anti-Hermitian matrix
	\begin{equation}
	\psi^{(0)}_{B}=-i\frac{g^2 \gamma_{P}}{2}
	\begin{pmatrix}
	1 & -1 \\
	-1 & 1 \\
	\end{pmatrix}.
	\label{eq:k0}
	\end{equation}
The representation (\ref{eq:k0}) is given with respect to the basis consisting of $\dket{0;P}$ and $\dket{0;-P}$.
The right-eigenvalue equation
	\begin{equation}
	\psi^{(0)}_{B}\dket{\phi^{(0)}_{\alpha;P}}=z^{(0)}_{\alpha;P}\dket{\phi^{(0)}_{\alpha;P}},
	\end{equation}
has a solution consisting of the eigenvalues and eigenvectors,
	\begin{subequations}
	\begin{equation}
	z^{(0)}_{+;P}=0,
	\label{eq:right-eigenvalue-k0-p}
	\end{equation}
with
	\begin{equation}
    \dket{\phi^{(0)}_{+;P}}=\frac{1}{\sqrt{2}}(\dket{0;|P|}+\dket{0;-|P|}),
    \label{eq:right-eigenstate-k0-p}
	\end{equation}
	\end{subequations}
and
	\begin{subequations}
	\begin{equation}
	z^{(0)}_{-:P}=-ig^2\gamma_{P},
	\label{eq:right-eigenvalue-k0-n}
	\end{equation}
with
	\begin{equation}
    \dket{\phi^{(0)}_{-;P}}\!=\!\frac{1}{\sqrt{2}}(\dket{0;|P|}-\dket{0;-|P|}).
	\label{eq:right-eigenstate-k0-n}
	\end{equation}
	\end{subequations}
Here we note that $\dket{\phi^{(0)}_{\pm;P}}=\dket{\phi^{(0)}_{\pm;-P}}$ and 
$z^{(0)}_{\pm;P}=z^{(0)}_{\pm;-P}$. 
The eigenstate $\dket{\phi^{(0)}_{+;P}}$ is the equilibrium mode,
while the eigenstate $\dket{\phi^{(0)}_{-;P}}$ is the decay mode of the momentum distribution function.
As a result, the momentum relaxation time $\tau_{P}$ is given by
	\begin{equation}
	\tau_{P}=\frac{1}{g^2\gamma_{P}}.
	\end{equation}
Since the operator (\ref{eq:k0}) is a symmetric anti-Hermitian matrix, each left-eigenvector is the transpose (Hermitian conjugate) of the corresponding right-eigenvector (\ref{eq:right-eigenstate-k0-p}) and (\ref{eq:right-eigenstate-k0-n}),
	\begin{equation}
	\dbra{{\tilde\phi}^{(0)}_{+;P}}
        =\frac{1}{\sqrt{2}}(\dbra{0;|P|}+\dbra{0;-|P|})
        =\dbra{\phi^{(0)}_{+;P}},
	\end{equation}
	\begin{equation}
	\dbra{{\tilde\phi}^{(0)}_{-;P}}
        =\frac{1}{\sqrt{2}}(\dbra{0;|P|}-\dbra{0;-|P|})
        =\dbra{\phi^{(0)}_{-;P}}.
	\end{equation}
These eigenstates satisfy the bi-orthonormality,
	\begin{equation}
	\dbraket{{\tilde\phi}^{(0)}_{\alpha;P}}{\phi^{(0)}_{\alpha';P'}}
        =\delta_{\alpha;\alpha'}\left[\delta(P-P')+\delta(P+P')\right],
	\end{equation}
and the bi-completeness relation,
	\begin{equation}
	\sum_{\alpha=\pm}\int^{\infty}_{0}dP\dket{\phi^{(0)}_{\alpha;P}}\!\dbra
        {{\tilde\phi}^{(0)}_{\alpha;P}}={\hat p}^{(0)},
	\end{equation}
where the integration over $P$ is restricted to the region $P\geq 0$ to avoid double counting.

\subsection{The spatially inhomogeneous situation: $k\not=0$}

Let us now consider the spatially inhomogeneous situation with $k\not=0$.
In this situation, the eigenvalue equation of the collision operator (\ref{eq:2matrix-boltzmann}) is
	\begin{equation}
	{\rm det}[\psi^{(k)}_{B}-z{\hat I}_{2}]\!\!=\!\!\bigpar{z+i\frac{g^2\gamma_{P}}{2}}^{2}
        \!\!\!-\bigpar{\frac{kP}{m}}^{2}\!\!\!+\bigpar{\frac{g^2\gamma_{P}}{2}}^{2}\!\!\!=0,
	\label{eq:character}
	\end{equation}
where ${\hat I}_{2}$ is the unit matrix of size $2$.
Hence, the eigenvalues are
	\begin{equation}
	z^{(k)}_{\pm;P}=-i\frac{g^2\gamma_{P}}{2}\pm\frac{|P|}{m}(k^{2}-k_{P}^{2})^{1/2},
	\label{eq:eigenvalues}
	\end{equation}
where
	\begin{equation}
	k_{P}\equiv\frac{g^2\gamma_{P}}{2|P|/m}\equiv\frac{1}{l_{P}},
	\label{eq:kpdef}
	\end{equation}
is a wavenumber that is equal to the inverse of the mean-free-length $l_{P}$ 
of the test particle with momentum $P$. 
Here we choose the branch of the square root function as
	\begin{equation}
	(k^{2}-k_{P}^{2})^{1/2}\equiv
		\begin{cases}
		\sqrt{k^{2}-k_{P}^{2}}\hspace{18pt}(k\geq k_{P})\\
		i\sqrt{k_{P}^{2}-k^{2}}\hspace{13pt}(|k|< k_{P})\\
		-\sqrt{k^{2}-k_{P}^{2}}\hspace{10pt}(k\leq-k_{P})\hspace{5pt}.
		\end{cases}
	\end{equation}
These eigenvalues tend in the limit $|k|\rightarrow0$ as
	\begin{eqnarray}
	\begin{split}
	&z^{(k)}_{+;P}\rightarrow z^{(0)}_{+;P},\\
	&z^{(k)}_{-;P}\rightarrow z^{(0)}_{-;P},
	\end{split}
	\end{eqnarray}
with (\ref{eq:right-eigenvalue-k0-p}) and (\ref{eq:right-eigenvalue-k0-n}).

For $|k|/k_{P}\ll1$, one can expand the eigenvalue $z^{(k)}_{+;P}$ and $z^{(k)}_{-;P}$ as
\begin{subequations}
\begin{equation}
z^{(k)}_{+;P}=-i\frac{g^2\gamma_{P}}{4}\bigpar{\frac{k}{k_{P}}}^{2}-i\frac{g^2\gamma_{P}}{16}\bigpar{\frac{k}{k_{P}}}^{4}+O\bigpar{\frac{k}{k_{P}}}^{6},
\label{eq:expansion_hydrodynamics}
\end{equation}
and
\begin{eqnarray}
&&z^{(k)}_{-;P}=-ig^2\gamma_{P}+i\frac{g^2\gamma_{P}}{4}\bigpar{\frac{k}{k_{P}}}^{2}+i\frac{g^2\gamma_{P}}{16}\bigpar{\frac{k}{k_{P}}}^{4}\nonumber\\
&&\hspace{30pt}+O\bigpar{\frac{k}{k_{P}}}^{6},
\label{eq:expansion_hydrodynamics-2}
\end{eqnarray}
\end{subequations}
First few terms of the expansion (\ref{eq:expansion_hydrodynamics}) give transport coefficients of hydrodynamic equations.
For instance, the first term, which is second order in $k$, gives the diffusion coefficient,
\begin{equation}
D_{P}\equiv\frac{g^2\gamma_{P}}{4k_{P}^{2}}=\frac{(P/m)^{2}}{g^2\gamma_{P}}.
\label{eq:diffusion-coeff}
\end{equation}
We have $D_{P}\to 0$ for $P\to 0$ (see (\ref{eq:Vq_condition})).
The higher order terms of the expansion also give transport coefficients of the Burnett equation (see e.g. \cite{resibois}).

\begin{figure}[t]
	\begin{center}
	\includegraphics[width=0.9 \linewidth]{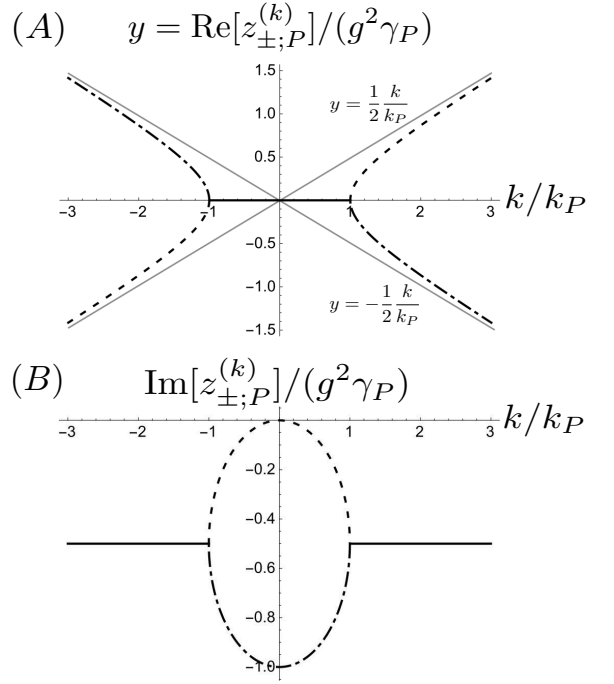}
	\end{center}
	\vspace{-2em}
	\caption{Eigenvalues of the Boltzmann collision operator (\ref{eq:eigenvalues}) are drawn as functions of $k$. (A) is the real part and (B) is imaginary part. In each figure, the dashed lines represent eigenvalue with $\alpha=+$ and the dot-dashed lines represent eigenvalue with $\alpha=-$. The solid lines represent that these two lines are overlapping. The gray lines in (A) represents eigenvalues of the Liouvillian for a free light-mass particle $y=\pm (1/2)(k/k_{P})$.}
	\label{fig1} 
	\end{figure}

In FIG.\ref{fig1}, we show a $k$-dependence of the real part and the imaginary part of the eigenvalues for $P\not=0$.
In the figures, the dashed lines and the dot-dashed lines represent the eigenvalues $z^{(k)}_{+;P}$ and $z^{(k)}_{-;P}$, respectively.
The solid lines represent that these two lines overlap.
Note that the eigenvelues (\ref{eq:eigenvalues}) can be rewritten as,
\begin{equation}\label{eq:scaled_eigenvalues}
\frac{z_{\pm;P}^{(k)}}{g^2 \gamma_P}
=
-\frac{i}{2} \pm \frac{1}{2}\left[\left(\frac{k}{k_P}\right)^2 - 1 \right]^{1/2}.
\end{equation}
Hence the eigenvalues depend on $P$ only through the $P$-dependences of $g^2 \gamma_P$ and $k_P$.
As a result, the spectrum is universal in the sense that
the eigenvalues are independent of $P$, 
if we use the $P$-dependent units in which $g^2 \gamma_P=1$ and $k_P=1$.

The eigenstates corresponding to the eigenvalues in (\ref{eq:eigenvalues}) are
	\begin{subequations}
	\begin{eqnarray}
	\begin{split}
	&\dket{\chi^{(k)}_{\pm;P}}=\Biggr[1\pm\frac{(k^{2}-k_{P}^{2})_{+}^{1/2}}{k}\Biggr]^{1/2}\dket{k;|P|}\\
	&\hspace{35pt}+i\frac{|k|}{k}\Biggr[1\mp\frac{(k^{2}-k_{P}^{2})^{1/2}}{k}\Biggr]^{1/2}\dket{k;-|P|},
	\end{split}
	\end{eqnarray}
	\begin{eqnarray}
	\begin{split}
	&\dbra{{\tilde\chi}^{(k)}_{\pm;P}}=\frac{|k|}{k}\Biggr[1\pm\frac{(k^{2}-k_{P}^{2})^{1/2}}{k}\Biggr]^{1/2}\dbra{k;|P|}\\
	&\hspace{35pt}+i\Biggr[1\mp\frac{(k^{2}-k_{P}^{2})^{1/2}}{k}\Biggr]^{1/2}\dbra{k;-|P|},
	\end{split}
	\end{eqnarray}
	\label{eq:eigenvectors}
	\end{subequations}
respectively. 
We note that $z^{(k)}_{\pm;P}=z^{(k)}_{\pm;-P}$ and $\dket{\chi^{(k)}_{\pm;P}}=\dket{\chi^{(k)}_{\pm;-P}}$.
Moreover, we note that $\dbra{{\tilde\chi}^{(k)}_{\pm;P}}$ is not the Hermitian conjugate of $\dket{\chi^{(k)}_{\pm;P}}$, 
because $\psi^{(k)}_B$ with $k\not=0$ is neither a Hermitian operator nor an anti-Hermitian operator.

In (\ref{eq:eigenvectors}), we have not yet normalized the eigenvectors 
by taking account of the fact that we have a diverging normalization constant at $k=\pm k_{P}$ 
(see (\ref{eq:normalization})).
The inner products of these right- and left-eigenstates are given by
	\begin{equation}
	\dbraket{{\tilde\chi}^{(k)}_{\pm;P}}{\chi^{(k)}_{\pm;P'}}
        =\pm\frac{2(k^{2}-k_{P}^{2})^{1/2}}{|k|}\left[\delta(P-P')+\delta(P+P')\right].
	\label{eq:inner-prod}
	\end{equation}
Then, normalized eigenstates for $k\not=\pm k_{P}$ are given by
	\begin{eqnarray}
	\begin{split}
	&\dket{\phi^{(k)}_{\pm;P}}\equiv\sqrt{N^{(k)}_{\pm;P}}\dket{\chi^{(k)}_{\pm;P}},\\
	&\dbra{{\tilde\phi}^{(k)}_{\pm;P}}\equiv\sqrt{N^{(k)}_{\pm;P}}\dbra{{\tilde\chi}^{(k)}_{\pm;P}},
	\label{eq:neigenvectors}
	\end{split}
	\end{eqnarray}
where the normalization constants are
	\begin{equation}
	N^{(k)}_{\pm;P}\equiv\pm\frac{|k|}{2(k^{2}-k_{P}^{2})^{1/2}}.
	\label{eq:normalization}
	\end{equation}
For $k\not=\pm k_{P}$, they satisfy following bi-orthonormal and bi-completeness relations,
	\begin{equation}
	\dbraket{{\tilde\phi}^{(k)}_{\alpha;P}}{\phi^{(k)}_{\alpha';P'}}=\delta_{\alpha,\alpha'}
        \left[\delta(P-P')+\delta(P+P')\right],
	\end{equation}
	\begin{equation}
	\sum_{\alpha=\pm}\int^{\infty}_{0}dP\dket{\phi^{(k)}_{\alpha;P}}\!\dbra{{\tilde\phi}^{(k)}_{\alpha;P}}={\hat p}^{(k)}.
	\label{eq:bi-comp-k}
	\end{equation}

The points $k=\pm k_{P}$ in Eq.(\ref{eq:eigenvalues}) are branch points of the eigenvalues in the complex $k$ plane.
At these points, both eigenvalues and eigenvectors coalesce as, for $k=k_{P}$,
	\begin{subequations}
	\begin{equation}
	z^{(k_{P})}_{\pm;P}=-i\frac{g^2\gamma_{P}}{2},
	\end{equation}
	\begin{equation}
	\dket{\chi^{(k_{P})}_{\pm;P}}=\dket{k_{P};|P|}+i\dket{k_{P};-|P|},
	\end{equation}
	\begin{equation}
	\dbra{{\tilde{\chi}}^{(k_{P})}_{\pm;P}}=\dbra{k_{P};|P|}+i\dbra{k_{P};-|P|},
	\end{equation}
	\end{subequations}
and as, for $k=-k_{P}$,
	\begin{subequations}
	\begin{equation}
	z^{(-k_{P})}_{\pm;P}=-i\frac{g^2\gamma_{P}}{2},
	\end{equation}
	\begin{equation}
	\dket{\chi^{(-k_{P})}_{\pm;P}}=\dket{-k_{P};|P|}-i\dket{-k_{P};-|P|},
	\end{equation}
	\begin{equation}
	\dbra{{\tilde{\chi}}^{(-k_{P})}_{\pm;P}}=-\dbra{-k_{P};|P|}+i\dbra{-k_{P};-|P|}.
	\end{equation}
	\end{subequations}

Since there is only one linearly independent eigenvector at the points, 
the Boltzmann collision operator (\ref{eq:2matrix-boltzmann}) can not be diagonalized.
Instead, the collision operator has the Jordan block structure at these points (see \cite{bhamathi,hashimoto2015}).
This coalescence of eigenvectors does not take place at the usual degeneracy point 
of the eigenvalues of a Hermitian operator for which a degenerate eigenvalue is shared by two distinct eigenstates.
Such a branch point of the spectrum of a non-Hermitian operator 
in the parameter space are called an {\it exceptional point} (EP) \cite{kato66}, 
which is also known as a {\it non-Hermitian degeneracy points} \cite{berry04}.
In the previous paper \cite{hashimoto2015}, we have introduced a divergence free representation at EPs 
by continuously extending the Jordan block representation away from EPs.


Note that the imaginary part of the eigenvalues $z^{(k)} _{\pm;P}$ 
remain constant $-g^2\gamma_{P}/2$ for $|k|\ge k_{P}$, and they do not vanish in the limit $|k| \to \infty$. 
As a result, one cannot get any information of the microscopic space structure 
of the interaction through the Boltzmann collision operator.  
This is clearly a limitation of the Boltzmann approximation 
to analyze microscopic origin of irreversibility in dynamical processes.


\section{Time evolution of the Wigner distribution function and the exceptional point}

In this section, we discuss the relation of the spectral structure to the transport processes in the system.
In the previous paper \cite{hashimoto2015}, we have shown that the exceptional points 
in the spectrum of the Boltzmann collision operator lead to the telegraph equation 
in the time evolution in coordinate space.
Here, we study further detail of the time evolution.
The results presented in this section are summarized as follows:\\
We found shifting motion of the peak of the distribution function in space 
in addition to spreading as a result of a diffusion type process.  
However, the mechanism of the shifting motion is very different 
in the domain $|k|\le k_{P}$ from that in the domain $|k| > k_{P}$.
\begin{enumerate}
\item
For $|k|\le k_{P}$, the shifting motion comes from asymmetry in the momentum distribution 
before the momentum relaxation is complete.

\item
For $|k| > k_{P}$, the shifting motion comes from the real part of the eigenvalue that leads to a wave propagation 
with the initial velocity $P/m$.

\end{enumerate}

To see this, we shall analyze time evolution of the Wigner distribution function for the test particle
	\begin{equation}
	f^{W}(X,P,t)=\int_{-\infty}^{\infty}dk\;f_{k}(P,t)e^{ikX},
	\end{equation}
with
	\begin{equation}
	f_{k}(P,t)\equiv\dbraket{k;P}{f(t)}.
	\end{equation}
Multiplying $\dbra{k;P}$ and using the completeness relation (\ref{eq:bi-comp-k}) to the formal solution of the Boltzmann equation
	\begin{equation}
	{\hat p}^{(k)}\dket{f(t)}=e^{-i\psi^{(k)}_{B}t}{\hat p}^{(k)}\dket{f(0)},
	\end{equation}
we have the following expression of the Fouier transformation of the Wigner distribution function $f_{k}(P,t)$
	\begin{eqnarray}
	\begin{split}
	f_{k}(P,t)&=\frac{1}{2}\Bigr(e^{-iz^{(k)}_{+;P}t}+e^{-iz^{(k)}_{-;P}t}\Bigr)f_{k}(P,0)\\
	&\hspace{5pt}+\frac{kP}{m}\Biggr(\frac{e^{-iz^{(k)}_{+;P}t}-e^{-iz^{(k)}_{-;P}t}}{z^{(k)}_{+;P}-z^{(k)}_{-;P}}\Biggr)f_{k}(P,0)\\
	&\hspace{5pt}+i\frac{g^{2}\gamma_{P}}{2}\Biggr(\frac{e^{-iz^{(k)}_{+;P}t}-e^{-iz^{(k)}_{-;P}t}}{z^{(k)}_{+;P}-z^{(k)}_{-;P}}\Biggr)f_{k}(-P,0).
	\end{split}
	\label{eq:Fourier-WDF}
	\end{eqnarray}
The first two terms correspond to time evolution due to forward scattering and the third term corresponds to time evolution due to backward scattering.

Note that the expression in (\ref{eq:Fourier-WDF}) is regular at the EPs where $z^{(k)}_{+;P}=z^{(k)}_{-;P}$.
Indeed, by taking the limit $k\rightarrow k_{P}$ or $k\rightarrow-k_{P}$, we have
	\begin{eqnarray}
	\begin{split}
	f_{\pm k_{P}}(P,t)&=e^{-\frac{g^2\gamma_{P}}{2}t}f_{\pm k_{P}}(P,0)\\
	&\hspace{5pt}\mp i\frac{k_{P}P}{m}te^{-\frac{g^2\gamma_{P}}{2}t}f_{\pm k_{P}}(P,0)\\
	&\hspace{5pt}+\frac{g^2\gamma_{P}}{2}te^{-\frac{g^2\gamma_{P}}{2}t}f_{\pm k_{P}}(-P,0),
	\end{split}
	\label{eq:critical}
	\end{eqnarray}
which is identical with the expression derived in terms of the Jordan block representation 
presented in \cite{hashimoto2015}.

In the second and the third terms in (\ref{eq:critical}), we have critical damping factors $t\exp[-(g^2\gamma_{P}/2)t]$.
Since the eigenvalues $z^{(k)}_{+;P}$ and $z^{(k)}_{-;P}$ change their values from pure imaginary values 
for $|k|<k_{P}$ to complex values for $|k|>k_{P}$, 
the time evolution of the Fourier component (\ref{eq:Fourier-WDF}) changes 
from an over-damping to a damped oscillation as a critical damping behavior at the EPs.

In order to see the relation of the spectral structure to the time evolution of the Wigner distribution function, we consider the following situation as an initial condition,
	\begin{equation}
	f_{k}(P,0)=\chi_{k_{b}}(k)h(P),
	\label{eq:initial}
	\end{equation}
where $\chi_{k_{b}}(k)$ is a step function which is defined with a given value of $k_{b}$ by
	\begin{equation}
	\chi_{k_{b}}(k)=
	\begin{cases}
	1\hspace{10pt}(|k|\leq k_{b}),\\
	0\hspace{10pt}(|k|>k_{b}),
	\end{cases}
	\end{equation}
and $h(P)$ is a momentum distribution function that is normalized as
	\begin{equation}
	\int^{\infty}_{-\infty}dP\; h(P)=1.
	\end{equation}
To extract the essence of the mechanism of the shifting motion, we here assume
	\begin{equation}
	h(P)=0\;\;\;{\rm for}\;\;\; P<0,
	\end{equation}
i.e., initial distribution is composed of particles with positive momentum.


We use the units in which $l_{P}=1/k_{P}=1$ and $\tau_{P}=1/(g^{2}\gamma_{P})=1$, 
when we present results of numerical calculations (see e.g. FIG.\ref{fig2}).
With these units, the eigenvalues and the eigenvectors are independent of the value of $P$ (see (\ref{eq:scaled_eigenvalues}) and (\ref{eq:eigenvectors})).
Therefore the dynamics described by the Boltzmann equation is universal 
in the sense that the time evolution equation in the units is the same regardless of the actual value of $P$.

\subsection{Time evolution with the spectrum in $|k|\leq k_{P}$}\label{sec:6A}

Let us first consider the situation where the initial distribution is composed of the Fourier components with $k$ in the region $|k|\leq k_{P}$.
Hence, we take
	\begin{equation}
	k_{b}\leq k_{P}.
	\end{equation}
In this case, the time evolution of the Wigner distribution function for $P>0$ is expressed by
	\begin{eqnarray}
	\begin{split}
	f^{W}(X,P,t)&=\int^{k_{b}}_{-k_{b}}dkf_{k}(P,t)e^{ikX}\\
	&=\int^{k_{b}}_{-k_{b}}dk\frac{k_{P}}{2\sqrt{k_{P}^{2}-k^{2}}}\\
	&\hspace{5pt}\times\biggl[e^{-iz^{(k)}_{+;P}t}\cos(kX-\varphi_{k,P})\\
	&\hspace{12pt}+e^{-iz^{(k)}_{-;P}t}\cos(kX+\varphi_{k,P})\biggr]f_{k}(P,0),
	\end{split}
	\label{eq:distribution1}
	\end{eqnarray}
where
	\begin{equation}
	\varphi_{k,P}\equiv\arctan\Biggr[\frac{k}{\sqrt{k_{P}^{2}-k^{2}}}\Biggr].
	\label{eq:phi}
	\end{equation}

\begin{figure}[t]
\begin{center}
\includegraphics[width=0.85 \linewidth]{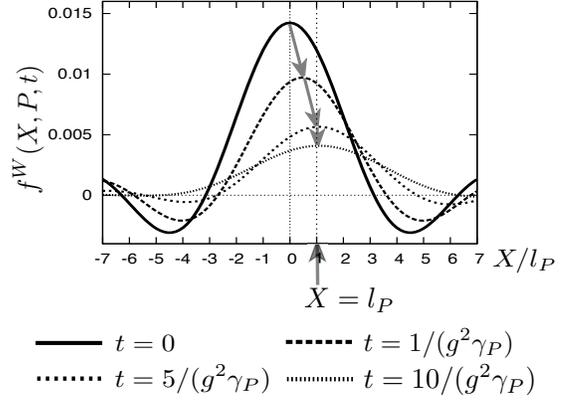}
\end{center}
\vspace{-2em}
\caption{Time evolution of the Wigner distribution function for $P>0$. The solid line represents the initial distribution. The initial distribution is given by Eq.(\ref{eq:initial}) with $k_{b}=k_{P}$. The initial distribution evolves to the distributions represented by the dotted lines as represented by the arrows. This figure does not depend on the value of $P$.}
\label{fig2}
\end{figure}

In FIG.\ref{fig2}, we present the time evolution of Eq.(\ref{eq:distribution1}) in $X$ space.
In the figure, the solid line is the initial distribution, and the dashed lines are the distribution of later times.
Here, and in the following figures in this section, we have not specified the value of $P$, since the time evolutions presented in this figure and in the following figures are independent of $P$ in the units in which $l_{P}=1$ and $\tau_{P}=1$.

As shown in the figure, the peak of the distribution function shifts toward $X=l_{P}$ in the first stage of its time evolution.
Afterwards, it stops the shifting motion and spreads its width symmetrically around the position $X=l_{P}$, where the time evolution is described by the diffusion equation with the diffusion coefficient $D_{P}$ in (\ref{eq:diffusion-coeff}).
This can be seen by the fact that (\ref{eq:distribution1}) can be approximated 
after the above mentioned first stage as,
\begin{equation}
f^W(X,P,t)\simeq\int_{-\infty}^\infty dk \frac{1}{2} e^{-D_P k^2 t} \cos[k(X-l_P)]f_k(P,0)
\end{equation}
which is a solution of the diffusion equation with a diffusion coefficient $D_P$ given by (\ref{eq:diffusion-coeff}).

We note that, for $t\rightarrow\infty$, the momentum distribution function
	\begin{equation}
	f_0(P,t)\equiv\frac{1}{2\pi}\int_{-\infty}^\infty dXf^W(X,P,t),
	\end{equation}
is stationary as
	\begin{equation}
	f_0(P,t)=f_{0}(-P,t)=\frac{1}{2}f_{0}(P,0).
	\end{equation}
This implies that the momentum relaxation has been completed,
and this is the reason that the peak of the distribution no longer moves.

\subsubsection*{\underline{Mechanism of the shift for the component with $|k|\leq k_{P}$}}

\begin{figure}[t]
\begin{center}
\includegraphics[width=0.85 \linewidth]{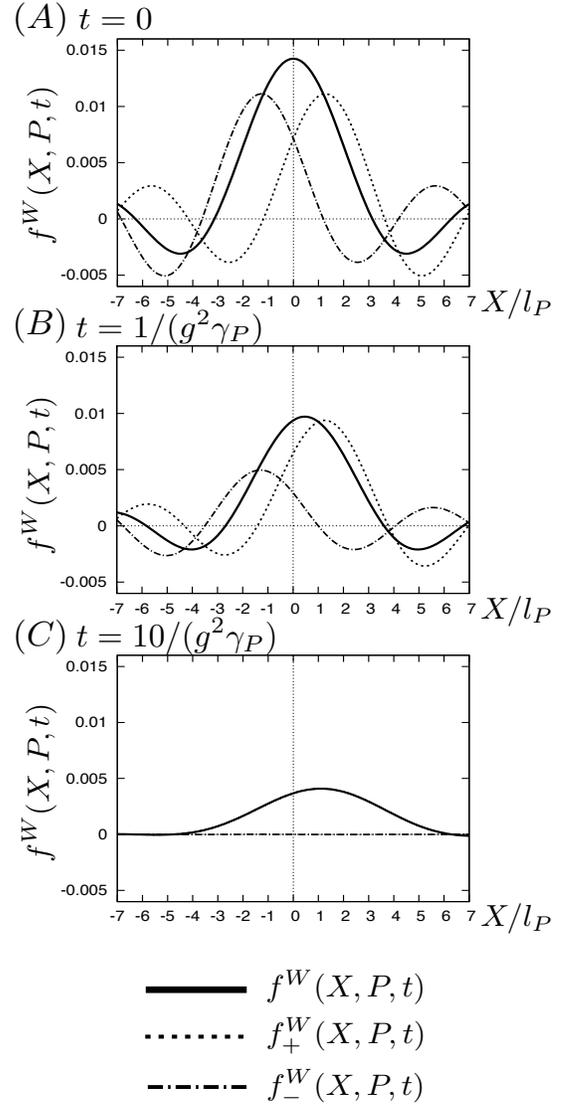}
\end{center}
\vspace{-2em}
\caption{Time evolution of the distribution function $f^{W}(X,P,t)$, its ``$+$'' component $f^{W}_{+}$ and its ``$-$'' component $f^{W}_{-}$ are presented for (A) $t=0$, (B) $t=1/(g^{2}\gamma_{P})$ and (C) $t=10/(g^{2}\gamma_{P})$. In each figure, the solid line represents $f^{W}$, the dashed line represents $f^{W}_{+}$ and the dot-dashed line represents $f^{W}_{-}$. The initial condition is given by Eq.(\ref{eq:initial}) with $k=k_{P}$. This figure does not depend on the value of $P$.}
\end{figure}

In the above, we have shown that the peak of the distribution function in $X$ space shifts in spite of the fact that there is no real component of the spectrum for $|k|\leq k_{P}$.
We now show that this is a result of the fact that the components of the distribution function correspond to $z^{(k)}_{+;P_{0}}$ and $z^{(k)}_{-;P_{0}}$ branches of the eigenvalues (\ref{eq:eigenvalues}) have asymmetry in $X$ space, and their decay rates are different as
	\begin{equation}
	|{\rm Im}[z^{(k)}_{+;P}]|\leq|{\rm Im}[z^{(k)}_{-;P}]|\hspace{8pt}{\rm for}\hspace{8pt}|k|\leq k_{P}.
	\label{eq:branchs}
	\end{equation}
(see also FIG.1(B))

To see this, let us first define the following two projection operators
	\begin{equation}
	{\hat p}^{(k)}_{\pm} \equiv \int_0^\infty dP\dket{\phi^{(k)}_{\pm;P}}\!\dbra{{\tilde\phi}^{(k)}_{\pm;P}}.
	\end{equation}
They satisfy
	\begin{equation}
	{\hat p}^{(k)}={\hat p}^{(k)}_{+}+{\hat p}^{(k)}_{-},
	\end{equation}
as in (\ref{eq:bi-comp-k}).

Time evolution of both ${\hat p}^{(k)}_{+}$ and ${\hat p}^{(k)}_{-}$ components of $\dket{f(t)}$ with an initial condition (\ref{eq:initial}) and with $k_{b}\leq k_{P}$ for $P>0$ are given by
	\begin{eqnarray}
	\begin{split}
	f^{W}_{\pm}(X,P,t)&\equiv\int^{\infty}_{-\infty}dk\dbra{k;P}{\hat p}^{(k)}_{\pm}\dket{f(t)}e^{ikX}\\
	&=\int^{k_{b}}_{-k_{b}}dk\frac{k_{P}}{2\sqrt{k_{P}^{2}-k^{2}}}\\
	&\hspace{5pt}\times e^{-iz^{(k)}_{\pm;P}t}\cos(kX\mp\varphi_{k,P})f_{k}(P,0).
	\end{split}
	\end{eqnarray}
Then, we have
	\begin{equation}
	f^{W}(X,P,t)=f^{W}_{+}(X,P,t)+f^{W}_{-}(X,P,t).
	\end{equation}

In FIG.3, we present the time evolution of $f^{W}_{+}$, $f^{W}_{-}$ and $f^{W}$ in $X$ space 
with a fixed value of momentum $P>0$.
In each figure in the FIG.3, the solid line represents $f^{W}$, the dashed line represents $f^{W}_{+}$ and the dot-dashed line represents $f^{W}_{-}$.
As shown in the figures, the position of the peak of $f^{W}_{+}$ is in the region $X>0$ and the position of the peak of $f^{W}_{-}$ is in $X<0$.
As time passes, the peak of $f^{W}_{-}$ decays more rapidly than the peak of $f^{W}_{+}$, and thus the peak of the total distribution $f^{W}$ shifts forward as shown in FIGS.3(B) and 3(C).

Physically, the shifting motion corresponds to center of mass motion of the distribution due to asymmetry of the initial momentum distribution function.

\subsection{Time evolution with the spectrum in $|k|>k_{P}$}

\begin{figure}[t]
\begin{center}
\includegraphics[width=0.85 \linewidth]{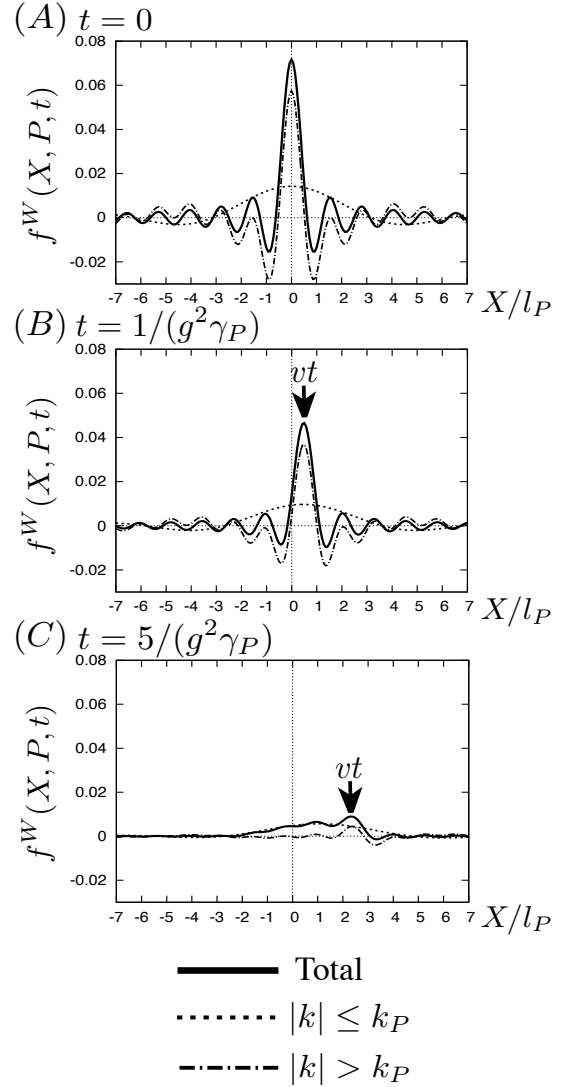}
\end{center}
\vspace{-2em}
\caption{Time evolution of the Wigner distribution function for $P>0$. The initial distribution (A) is given by (\ref{eq:initial}) with $k_{b}=5.0k_{P}$. The distribution functions are shown for (A) $t=0$, (B) $t=1/(g^{2}\gamma_{P})$ and (C) $t=5/(g^{2}\gamma_{P})$. In each figure, the solid line represents total value of the function $f^{W}$, the dashed line and the dot-dashed line represents $f^{W}_{{\rm d}}$ and $f^{W}_{{\rm p}}$, respectively. This figure does not depend on the value of $P$.}
\end{figure}

In Sec. V, we showed that the spectrum of the Liouvillian takes complex values in the region $|k|>k_{P}$ and their real part approaches to the eigenvalue for a free particle in the limit $|k|\rightarrow\infty$ (see FIG.1).
In this subsection, we discuss how this spectrum affects the time evolution of the system.
For this purpose, we analyze time evolution with an initial condition (\ref{eq:initial}) with
	\begin{equation}
	k_{b}>k_{P}.
	\end{equation}
For this initial condition, the time evolution of the Wigner distribution function for $P>0$ can be divided into two parts; 
(i) $f_\mathrm{d}^W$ with pure imaginary eigenvalues and (ii) $f_\mathrm{p}^W$ with complex eigenvalues as
	\begin{eqnarray}
	\begin{split}
	f^{W}(X,P,t)&\equiv\int^{\infty}_{-\infty}dkf_{k}(P,t)e^{ikX}\\
	&=f^{W}_{{\rm d}}(X,P,t)+f^{W}_{{\rm p}}(X,P,t),
	\end{split}
	\end{eqnarray}
where (i)
	\begin{eqnarray}
	\begin{split}
	f^{W}_{{\rm d}}(X,P,t)&\equiv\int^{k_{P}}_{-k_{P}}dkf_{k}(P,t)e^{ikX}\\
	&=\int^{k_{P}}_{-k_{P}}dk\frac{k_{P}}{\sqrt{k_{P}^{2}-k^{2}}}\\
	&\hspace{5pt}\times\biggl[e^{-iz^{(k)}_{+;P}t}\cos(kX-\varphi_{k,P})\\
	&\hspace{12pt}+e^{-iz^{(k)}_{-;P}t}\cos(kX+\varphi_{k,P})\biggr]f_{k}(P,0),
	\end{split}
	\label{eq:distribution3-1}
	\end{eqnarray}
and (ii)
	\begin{eqnarray}
	\begin{split}
	&f^{W}_{{\rm p}}(X,P,t)\equiv\biggr(\int^{-k_{P}}_{-k_{b}}+\int^{k_{b}}_{k_{P}}\biggr)dkf_{k}(P,t)e^{ikX}\\
	&=\frac{1}{2}\biggr(\int^{-k_{P}}_{-k_{b}}+\int^{k_{b}}_{k_{P}}\biggr)dk\; e^{-\frac{\gamma_{P}}{2}t}\\
	&\hspace{5pt}\times\biggl[e^{ik(X-\frac{P}{m}\sqrt{1-(k_{P}/k)^{2}}t)}\bigpar{1+\frac{1}{\sqrt{1-(k_{P}/k)^{2}}}}\\
	&\hspace{11pt}+e^{ik(X+\frac{P}{m}\sqrt{1-(k_{P}/k)^{2}}t)}\bigpar{1-\frac{1}{\sqrt{1-(k_{P}/k)^{2}}}}\biggr]\\
	&\hspace{5pt}\times f_{k}(P,0).
	\end{split}
	\label{eq:distribution3-2}
	\end{eqnarray}
Eq.(\ref{eq:distribution3-1}) consists of the over-damping components of (\ref{eq:Fourier-WDF}), 
while (\ref{eq:distribution3-2}) consists of the damped oscillation components.

In FIG.4, we show the time evolution of $f^{W}_{{\rm d}}$ and $f^{W}_{{\rm p}}$ as well as the total distribution function $f^{W}$ with a specific value of momentum $P>0$.
In each figure, the solid line represents the total distribution $f^{W}$, the dashed line represents $f^{W}_{{\rm d}}$ and the dot-dashed line represents $f^{W}_{{\rm p}}$.

The time evolution of $f^{W}_{{\rm d}}$ has the same character as the time evolution of (\ref{eq:distribution1}) 
discussed in the above subsection A, namely the distribution $f^{W}_{{\rm d}}(X,P,t)$ firstly shifts toward $X=l_{P}$, within the time interval $0\leq t\lesssim 1/(g^2\gamma_{P})$.
Afterwards, the distribution no longer shift 
and spreads its width by the diffusion process with its center fixed at $l_P$.
On the other hand, the distribution $f^{W}_{{\rm p}}(X,P,t)$ propagates as a wave packet 
with a velocity nearly equal to the initially give velocity $P/m$ and decays in time.
The wave propagation is due to the real part of the eigenvalues of the Liouvillian in the region $|k|>k_{P}$.


\section{Concluding remarks}

We have derived an irreversible kinetic equation for the weakly-coupled 1D quantum perfect Lorentz gas through the spectral analysis of the Liouvillian starting from the first principle of physics without relying upon phenomenological arguments. As demonstrated in this paper, irreversibility is an intrinsic dynamical property of the basic laws of physics coming from the resonance phenomena that is associated to a complex eigenvalue of the Liouvillian (which corresponds to a pole of the resolvent operator for the Liouvillian in the complex frequency space). In other words, the irreversibility is not an approximate concept coming from the human's limitations on controlling microscopic information of large systems as often stated in many textbooks in statistical mechanics (see e.g. \cite{Landau}).

Then, we have discussed a limitation of the applicability of the phenomenological Boltzmann equation.
As shown in this paper, one cannot analyze the irreversible processes by the Boltzmann equation 
in so small a scale where the microscopic structure of the interaction between the particles starts to play a role, because detailed space structure of the interaction is smeared out in the Boltzmann approximation. In order to describe the irreversible processes in such a small scale, we have to solve the nonlinear eigenvalue problem for the collision operator (\ref{eq:collop-pl}) beyond the Boltzmann approximation.

Moreover, we have shown that the perfect Lorentz gas with $m/M\rightarrow0$ cannot deal with the temperature dependence.
Hence, it is worth remarking that, in this approximation, one cannot study temperature dependent phenomena such as a transport process in the non-equilibrium steady state.
In order to deal with the temperature dependence, we need to go beyond the lowest order approximation in $m/M$ expansion.

We have also discussed an interesting shifting motion of the distribution function that spreads in moderately small spatial scale where hydrodynamic approximation is not applicable. There, we found two completely different mechanisms of the shifting motion; one is due to different value of the imaginary part of the eigenvalue of the Liouvillian, and dominates in a moderately small wavenumber associated to the inhomogeneity, while the other is due to the wave propagation associated to the real part in the eigenvalue that appears only in a moderately large wavenumber. 

In this paper we have not yet presented the solution of the nonlinear equation for the eigenvalue problem of the collision operator.  Only by solving the nonlinear problem we can describe the irreversible processes occurring in microscopic scale.  The precise limitation of the Boltzmann approximation is also analyzable only after we solve the nonlinear problem. We will show the solution of the nonlinear problem in a forthcoming paper where we will use the completeness relation of the eigen-basis of the Boltzmann equation that is presented in this paper.

\section{Acknowledgements}
We would like to express our sincere gratitude to Prof. H. Hayakawa for motivating us to launch this study and for his valuable comments.
We also thank to Dr. S. Garmon for fruitful discussions.
This work was supported by JSPS KAKENHI Grand Number 24540411.

\appendix*

\section{Temperature dependence}\label{app:eqd-collop-perfect}

In this appendix, we give an expression for the correction to the collision operator of the perfect Lorentz gas.

We expand the propagator in (\ref{eq:matrix-element-1}) as a power series of the ratio $m/M$,
	\begin{eqnarray}
	&&\frac{1}{z-(k-l)P/m-lP_{j}/M}\nonumber\\
	&&=\sum_{n=0}^\infty \frac{(lP_j/m)^n}{[z-(k-l)P/m]^{n+1}}\left(\frac{m}{M}\right)^n.
	\label{eq:prop-expans}
	\end{eqnarray}
For each order term, 
one can perform the integration over momentum of $N$ heavy particles 
using the following formulae for the Gaussian integrals,
	\begin{eqnarray}
	&&\int^{\infty}_{-\infty}\!\!dP_{i}^{N}P_{j}^{2n}\bigpar{\frac{1}{2\pi Mk_{B}T}}^{N/2}\!\prod_{i'=1}^{N}\exp\bigpar{-\frac{P_{i'}^{2}}{2Mk_{B}T}}\nonumber\\
	&&=I\times\!\int^{\infty}_{-\infty}\!\!dP_{j}P_{j}^{2n}\bigpar{\frac{1}{2\pi Mk_{B}T}}^{1/2}\!\!\!\!\!\exp\bigpar{-\frac{P_{j}^{2}}{2Mk_{B}T}}\nonumber\\
	&&=(2n-1)!!\bigpar{\frac{Mk_{B}T}{m}}^{n}m^{n},
	\label{eq:even-power}
	\end{eqnarray}
where $\int^{\infty}_{-\infty}dP^{N}_{i}$ stands for the integration over momenta of $N$ heavy particles, and
	\begin{eqnarray}
	&&\int^{\infty}_{-\infty}\!\!dP_{i}^{N}P_{j}^{2n+1}\bigpar{\frac{1}{2\pi Mk_{B}T}}^{N/2}\!\prod^{N}_{i'=1}\exp\bigpar{-\frac{P_{i'}^{2}}{2Mk_{B}T}}\nonumber\\
	&&=I\times\!\int^{\infty}_{-\infty}\!\!dP_{j}P_{j}^{2n+1}\bigpar{\frac{1}{2\pi Mk_{B}T}}^{1/2}\!\!\!\!\!\exp\bigpar{-\frac{P_{j}^{2}}{2Mk_{B}T}}\nonumber\\
	&&=0,
	\label{eq:odd-power}
	\end{eqnarray}
with
	\begin{eqnarray}
	I\equiv&&\prod^{N}_{i(\not=j)}\int^{\infty}_{-\infty}dP_{i}\bigpar{\frac{1}{2\pi Mk_{B}T}}^{1/2}\exp\bigpar{-\frac{P_{i}^{2}}{2Mk_{B}T}}\nonumber\\
	=&&1,
	\end{eqnarray}
which is a result of the normalization.
Note that the $T$ dependence of (\ref{eq:matrix-element-1}) comes from even power in the series expansion with $(m/M)$ in (\ref{eq:prop-expans}).
Hence, we
	\begin{eqnarray}
	&&\int^{\infty}_{-\infty}dP_{i}^{N}\frac{(lP_{j}/m)^{2n}}{[z-(k-l)P/m]^{2n+1}}\bigpar{\frac{m}{M}}^{2n}\rho^{eq}_{{\rm hev.}}\nonumber\\
	&&=\frac{(2n-1)!!}{[z-(k-l)P/m]^{2n+1}}\bigpar{\frac{l^{2}k_{B}T}{m}}^{n}\bigpar{\frac{m}{M}}^{n},
	\end{eqnarray}
for the integrals in (\ref{eq:matrix-element-1}) over the momenta of the heavy particles.
As a result, $T$ dependence appears already in the first order correction in $m/M$ in spite that (\ref{eq:matrix-element-1}) has a contribution in its even power.
Then, we have (\ref{eq:collop-pl}).

\end{document}